\documentclass[usenatbib]{mn2e}

\usepackage{graphicx}
\usepackage{times}

\newcommand{\beq}{\begin{equation}}
\newcommand{\eeq}{\end{equation}}
\newcommand{\beqarray}{\begin{eqnarray}}
\newcommand{\eeqarray}{\end{eqnarray}}
\def\lsim{\raise0.3ex\hbox{$\;<$\kern-0.75em\raise-1.1ex\hbox{$\sim\;$}}}
\def\gsim{\raise0.3ex\hbox{$\;>$\kern-0.75em\raise-1.1ex\hbox{$\sim\;$}}}

\usepackage[usenames]{color}

\title
[$p\gamma$ interactions in Galactic jets] 
{$p\gamma$ interactions in Galactic jets as a plausible origin of the positron excess} 

\author[Nayantara Gupta and Diego F. Torres]
{Nayantara Gupta $^1$ \thanks{nayan@rri.res.in} and Diego F. Torres $^2$ \thanks
{dtorres@ieec.uab.es}\\
$^1$ Astronomy \& Astrophysics Group, Raman Research Institute, C.V. Raman Avenue, Bangalore, India\\
$^2$ ICREA \& Institute of Space Sciences (IEEC-CSIC), Campus UAB, Torre C5, 2a planta, 08193 Barcelona, Spain}

\begin{document} 

\maketitle

\label{firstpage}

\begin{abstract}
The positron flux measured near Earth shows a rise with energy beyond 30 GeV. We show that this rise may be compatible with the production of positrons in $p  \gamma$ interactions in the jets of microquasars. 
 
\end{abstract}

\section{Introduction}

The electron and positron spectra measured by cosmic ray experiments 
(Boezio et al. 2000, DuVernois et al. 2001, Aguilar et al. 2013, Adriani et al. 2013, Ackermann et al. 2012) are interesting probes of Nature. 
Electrons and positrons are injected in our Galaxy by cosmic-ray sources and interactions during the propagation of cosmic ray protons and nuclei.
The electron flux times the electron's energy cube (e.g., in units of GeV$^2$ m$^{-2}$ s$^{-1}$ sr$^{-1}$) is a useful representation
{\bf of the spectrum}. It is nearly constant in the energy range  $10 $ GeV -- $1 $ TeV, and steeply falls down at higher energies.
The flux measured up to 4 TeV is composed of primary electrons and secondaries from various interactions.
Production of pairs in cosmic ray interactions with matter and radiation, electrons emitting pairs in a magnetic field $(e \, B\rightarrow e^+ \ e^-)$, and 
$\gamma \, \gamma\rightarrow e^+ \ e^-$ interactions contribute to the electron and positron fluxes equally. 
However, the recent measurements of the positron flux shows a rise beyond 30 GeV, which could have its origin either in conventional astrophysical sources or in dark matter annihilations, see e.g., Finkbeiner (2011), Gaggero et al. (2013), Cholis \& Hooper (2013), Mauro et al. (2014), Mertsch \& Sarkar (2014).    
A fraction of the positron flux measured up to 350 GeV originates from the 
pair-producing interactions mentioned above. Photo-hadronic interactions leading to the production of positrons $(p \,\gamma\rightarrow \Delta^+\rightarrow \pi^+ \, n$,
$\pi^+  \rightarrow  e^+ \, \nu_{\mu} \, \bar\nu_{\mu} \, \nu_e )$ may also contribute to the observed positron spectrum; see, e.g., Gupta \& Zhang (2008).

In the present work, we discuss the spectrum of positrons expected from these interactions in the jets of microquasars and how this excess flux may help explaining the recent measurements by PAMELA (Aguilar et al. 2013), AMS02 (Adriani et al. 2013) and Fermi experiments (Ackermann et al. 2012).
 Microquasars were considered earlier to explain the positron annihilation radiation at 511 keV (Guessoum, Jean \& Prantzos 2006, Vila \& Romero 2010).
The possibility of explaining the observed positron excess at tens of GeV with microquasars has been mentioned in the review by Fan et al. 2010, but up to now, no computation that demonstrates that this is indeed feasible nor any precision on which process would make this excess happen is available in literature. Here we use the positron data to study this possibility quantitatively.
   


The aim of this paper is not to provide a model of the inner MQ engine (and thus to construct a detailed SED of the photons emitted by it), which has been done in several papers, with different levels of detail. There are several situations where the $p\gamma$ process dominates over leptonic and other hadronic interactions (pp) at high energies. For instance see model A, C or D in the paper by Vila et al. 2012. In these models, the target radiation field considered for the $p\gamma$ interactions are the synchrotron photons of primary electrons and a detailed description of the location of the acceleration region and the magnetic field dependence along the jet axis
can be found. 

 \section{Photon normalization}


We shall denote the primary electron flux by $P_{\rm{e}^-}(E_{\rm{e}})$, whereas the secondary  electron flux produced in all pair producing interactions, including processes both local to the astrophysical sources and during the propagation of cosmic rays in Galaxy will be $F(E_{\rm{e}})$.
 
The injected fluxes of electrons and positrons are 
\begin{eqnarray}
Q_{\rm{e}^-}(E_{\rm{e}})=P_{\rm{e}^-}(E_{\rm{e}})+F(E_{\rm{e}}), \nonumber \\
Q_{\rm{e}^+}(E_{\rm{e}})=\phi_{p\gamma\rightarrow \nu e^+}(E_{\rm{e}})+F(E_{\rm{e}}).
\label{pos_inj}
\end{eqnarray}
Here, $\phi_{p\gamma\rightarrow \nu e^+}$ denotes the positron flux originating from the interaction channel $(p \,\gamma\rightarrow \Delta^+ \rightarrow 
\pi^+ \, n$, $\pi^+ \rightarrow  e^+ \, \nu_{\mu} \, \bar\nu_{\mu} \, \nu_e )$.
The electrons and positrons lose energy inside the sources before their escape. The terms in Eq.  (1) conceptually
denote the injected flux after including the losses inside the sources.

We parametrize the photon spectrum in the jet frame (in units of GeV$^{-1}$ cm$^{-3}$) 
of the $p\gamma$ sources with a broken power-law as
 \beq
\frac{dn_{\gamma}(\epsilon_{\rm{j}})}{d\epsilon_{\rm{j}}} 
=A \left\{ \begin{array}{l@{\quad \quad}l}
{\epsilon_{\rm{j}}}^{-\gamma_1} &
\epsilon_{\rm{j}}<\epsilon_{\rm{br,j}}\\{\epsilon_{\rm{br,j}}}^{\gamma_2-\gamma_1}
{\epsilon_{\rm{j}}}^{-\gamma_2} & \epsilon_{\rm{j}}>\epsilon_{\rm{br,j}}
\end{array}\right.
\label{phot_spec}
\eeq 
where the spectral indices of the photon spectrum are $\gamma_1$ and $\gamma_2$.  The break energy of the photon spectrum in the jet frame,
$\epsilon_{\rm{br,j}}$, is related to its value in the observer frame as
$\epsilon_{\rm{br}}=\delta_{\rm{D}}\epsilon_{\rm{br,j}}$.
The Doppler factor $\delta_{\rm{D}}$ is related to the Lorentz boost factor of the jet $\Gamma_{\rm{j}}$ and viewing angle in the observer's frame $\theta_{\rm{ob}}$ as
\beq
\delta_{\rm{D}}=\Gamma_{\rm{j}}^{-1}(1-\beta_{\rm{j}} \cos{\theta_{\rm{ob}}})^{-1}
\eeq
where $\beta_{\rm{j}}$ is the dimensionless speed of the jet frame with respect to the observer's frame on earth.
 The normalization constant $A$ can be calculated 
 from the energy density in the jet as
\beq
U=\int_{\epsilon_{\gamma,\rm{min}}}^{\epsilon_{\gamma,\rm{max}}} \epsilon_{\rm{j}}\frac{dn_{\gamma}(\epsilon_{\rm{j}})}{d\epsilon_{\rm{j}}} d\epsilon_{\rm{j}},
\eeq
where $\epsilon_{\rm{min,j}} \ll \epsilon_{\rm{br,j}} \ll \epsilon_{\rm{max},j}$. This gives 
\beq
A=\frac{U\epsilon_{\rm{br,j}}^{\gamma_1-2}}{[\frac{1}{\gamma_2-2}-\frac{1}{\gamma_1-2}]},
\eeq 
with $\gamma_i\neq 2,i=1,2$.

\section{$p \gamma$ interactions as a plausible source of the positron excess}

There are several channels for $p\gamma$ interactions, of which 
$p \, \gamma\rightarrow \Delta^+\rightarrow \pi^+ \, n$, 
$p \, \gamma\rightarrow \Delta^+ \rightarrow \pi^0 \, p$ 
are  dominant with nearly equal cross-section (Muecke et al. 2000, Kelner et al. 2008).
For lower energies, $p \,\gamma\rightarrow p \,e^+ \,e^-$ becomes the dominant channel.
The positively charged pions produced in $p \, \gamma\rightarrow \Delta^+\rightarrow \pi^+ \,n$ interactions decay to $\mu^+$ and $\nu_{\mu}$. 
Subsequently, the $\mu^+$ decay to $e^+$, $\bar\nu_{\mu}$ and $\nu_e$. Three neutrinos and a positron are produced at the final state. Thus, 
$p \gamma$ interactions  lead to asymmetry in the number of electrons and positrons in the Galaxy. Each $p \, \gamma$ interaction gives an extra positron.

The timescale for the  energy loss of a proton of energy
$E_{{\rm{p},j}}$ due to pion production, in the comoving frame, is (e.g., Waxman \& Bahcall 1997)
\beqarray
t_{\pi}^{-1}(E_{\rm{p,j}})=-\frac{1}{E_{\rm{p,j}}}\frac{dE_{\rm{p,j}}}{dt} 
\nonumber \\
=\frac{c}{2\Gamma_{\rm{p,j}}^2}\int_{\epsilon_{\rm{th}}}^{\infty}
d\epsilon \, \sigma_{\pi}(\epsilon)\xi(\epsilon)\epsilon 
\int_{\epsilon/2\Gamma_{\rm{p,j}}}^{\infty}
d\epsilon_{\rm{j}} \, {\epsilon_{\rm{j}}}^{-2} \frac{dn_{\gamma}(\epsilon_{\rm{j}})}{d\epsilon_{\rm{j}}} ,
\label{inv_time_pion}
\end{eqnarray}
where the Lorentz factor of a proton of energy $E_{\rm{p}}$ is $\Gamma_{\rm{p,j}}=E_{\rm{p,j}}/m_{\rm{p}} c^2$.
The threshold energy of pion production in the proton's rest frame is $\epsilon_{\rm{th}}=0.15$ GeV. 
The cross section $(\sigma_{\pi}(\epsilon))$ for $p  \gamma\rightarrow \Delta^+ \rightarrow \pi^+ \,n$ 
in the proton rest frame depends on the photon energy $\epsilon$ (in the proton rest frame too). 
At the resonance peak this cross section is $0.5$ mb. The average fractional energy ($\xi(\epsilon)$) of a proton going to a pion is 0.2. If this energy is equally shared by the four leptons at the final state then each positron takes $5\%$ of the parent proton's energy, 
$E_{\rm{e,j}}=0.05 E_{\rm{p,j}}$.

 The wind expansion time or dynamical timescale of the jet is 
$t_{\rm{dyn}}=R_{\rm{d}}/\delta_{\rm{D}} c$, where $R_{\rm{d}}$ denotes the radius of the emission region in the observer's frame.
The fractional proton energy lost to a pion during the dynamical timescale of the jet is 
$f_{\pi}(E_{\rm{p,j}})=t_{\rm{dyn}}/t_{\pi}(E_{\rm{p,j}})$ (Waxman \& Bahcall 1997). When the dynamical timescale is comparable to or greater than the pion production timescale, most of the protons will produce pions in $p\gamma$ interactions.
 One has to do the integration in Eq. (\ref{inv_time_pion}) to obtain the expression for $f_{\pi}(E_{\rm{p,j}})$. 
A simplified expression was given by Waxman \& Bahcall (1997),
whereas a more general expression involves the spectral indices $\gamma_1$ and $\gamma_2$ (Gupta \& Zhang 2007, Moharana \& Gupta 2012).

The break in the low energy photon spectrum ($\epsilon_{\rm{br}}$) reflects 
 the break energy in the proton spectrum ($E_{\rm{pb}}$),  due to the threshold energy condition of $p\gamma$ interactions
$
E_{pb} \epsilon_{\rm{br}}=0.14\, {\delta_{\rm{D}}}^2$ GeV$^2$.
The synchrotron luminosity, $L_{\gamma}$, is the product of the energy density of the photons and the volume of the emission region per unit time. If the emission is isotropic in the jet frame then
\beq
L_{\gamma}= 4 \, \pi \, {R_{\rm{d}}} ^2 \, {\delta_{\rm{D}}}^2 \, U \, c
\eeq
 For our purpose only the energy dependence of $f_{\pi}(E_{\rm{p}})$ is needed. 
 
In terms of the proton energy $E_{\rm{p}}$ and the break energy in the proton spectrum $E_{\rm{pb}}$, which appears due to the break in the low energy photon spectrum at $\epsilon_{\rm{br}}$ it is (Gupta \& Zhang 2007, Moharana \& Gupta 2012)
\beq
 f_{\pi}(E_{\rm{p}})\propto \left\{ \begin{array}{l@{\quad \quad}l}
(\frac{E_{\rm{p}}}{E_{\rm{pb}}})^{(\gamma_2-1)} ,   E_{\rm{p}}<E_{\rm{pb}}\\
(\frac{E_{\rm{p}}}{E_{\rm{pb}}})^{(\gamma_1-1)} ,  E_{\rm{p}}>E_{\rm{pb}}\end{array} \right. 
\label{fpi}
\eeq 
 The value of $E_{\rm{pb}}$ may be reflected in future measurements of the positron spectrum at higher energies.

As mentioned above, in $p\gamma$ interactions, the probabilities of $\pi^0$ and $\pi^+$ production are nearly equal. The fractional energy lost by a proton to a pion is $f_{\pi}(E_{\rm{p}})$. Assuming the energy of the $\pi^+$ is equally shared by the four leptons at the final state ($e^+$, $\nu_e$, $\nu_{\mu}$ and $\bar\nu_{\mu}$), one can write down the positron energy flux in terms of the luminosity in protons $L_{\rm{p}}$.
\beq
E_{\rm{e}}^2 \eta_{p\gamma\rightarrow \nu e^+}(E_{\rm{e}})=\frac{f_{\pi}(E_{\rm{p}})}{8}L_{\rm{p}}
\label{pos_enflux}
\eeq
The fraction $f_{\pi}(E_{\rm{p}})$ can also be written as $f_{\pi}(E_{\rm{e}})$ as it contains the ratio of $E_{\rm{p}}$ and $E_{\rm{pb}}$, and $E_{\rm{e}}=0.05E_{\rm{p}}$.
%
%
The flux of positrons per unit energy produced inside the source is  $\eta_{p\gamma\rightarrow \nu e^+}$ (GeV$^{-1}$ s$^{-1}$)$\propto E_{\rm{e}}^{(\gamma_2-3)}$ below the break energy. The spectral index $(\gamma_2-3)$ is obtained after dividing $f_{\pi}(E_{\rm{e}})\propto E_{\rm{e}}^{(\gamma_2-1)}$ by $E_{\rm{e}}^2$. 
 We are considering relativistic positrons with energy larger than 10 GeV, for 
which synchrotron and SSC losses are important.
If the average distance traversed by the electrons and positrons before losing energy is much less than the size of the production region
then the spectrum of the escaping positrons will be steeper by a factor of $1/E_{\rm{e}}$ (Berezinskii et al. 1990).  Thus the injected spectrum of positrons from the jets will have a spectrum, 
 $\phi_{p\gamma\rightarrow \nu e^+}(E_{\rm{e}})$(GeV$^{-1}$ s$^{-1}$)$\propto  E_{\rm{e}}^{(\gamma_2-4)}$.

\section{Diffused spectrum}

The diffusion equation gives the particle density as a function of space, time and energy for a known injection spectrum, diffusion coefficient and energy-loss term (Ginzburg \& Syrovatskii 1964, Berezinskii et al. 1990).
The energy dependent diffusion coefficient is
$
D(E)= D_0 E^{\delta},
$
where $\delta$ may vary between 0.33 (Kolmogorov model) and 0.6 (Kraichnan model) (Bernardo et al. 2010). Synchrotron and inverse Compton (IC) are the most important radiative loss mechanisms for the relativistic electrons and positrons propagating through the Galaxy.
 Figure 1. of \citet{jean} already shows that (even for a warm medium), electrons and positrons with an energy in excess of 30 GeV lose it mainly via synchrotron and IC losses. For GeV electrons and positrons, annihilation in ISM is not important at all and can be disregarded. 
  The energy loss term of electrons and positrons is 
\beq
\frac{dE}{dt}=b(E)=-\beta E^2,
\eeq
as both synchrotron and inverse Compton losses are proportional to $E^2$, $\beta=1.38\times 10^{-16}$~(GeV~\,s)$^{-1}$ (Volkov \& Lagutin 2013).
Their energy is then 
$
E(t)=E_0/(1+\beta E_0 t),
$
where $E=E_0$ at $t=0$.  
\par
The time taken for the electron or positron of initial energy $E_0$ to lose half of its energy ($E=E_0/2$) is 
\beq
\tau(E,E_0)=\int_{E_0}^{E} \frac{dE'}{b(E')} = \frac{1}{\beta E_0}.
\eeq
The average distance to which the particles have diffused while loosing energy is (Berezinskii et al. 1990)
\beq
\lambda(E,E_0)=\Big(\int_{E_0}^E \frac{D(E')dE'}{b(E')}\Big)^{1/2} .
\eeq
Thus, the distance traversed by a $30$ GeV positron before losing half of its energy is about $600$ pc in a medium with magnetic field of
5 $\mu$G and photon density of 1 eV/cm$^3$ assuming 
$D=10^{29}$ cm$^2$ s$^{-1}$.
We consider our Galaxy, with a radial extent of $R\sim 10$ kpc, thickness of the disc of $d\sim 150$ pc,
 and halo size of $h\sim 4 $ kpc. 
 If the sources are uniformly distributed in the Galactic disc, then for an injected spectrum $E^{-\alpha}$ 
 the propagated spectrum is proportional to $E^{-\alpha-1/2}$ for $\lambda(E)> d$ and $E^{-\alpha-1}$ for $\lambda(E) \ll d$ (Berezinskii et al. 1990). 
 Hence in our case ($\lambda(E)> d$) the propagated spectrum is expected to be $E^{-\alpha-1/2}$.
 
In reality, the LMXBs are distributed in the disc, bulge, and spheroid of the Milky Way (Bahcall \& Soneira 1980; Grimm, Gilfanov \& Sunyaev 2002; Revnivtsev et al. 2008) in an approximately 2:1:0.3 ratio. 
Thus, one would in principle need to use the distribution of sources from Bahcall \& Soneira 1980, also given in Eq.  (4--6) of Grimm, Gilfanov \& Sunyaev et al. 2002 to calculate the propagated spectrum of positrons with the transport equation given in Eq.  (5.6) of Berezinskii et al. 1990. 
Here, we have simplified this problem assuming a uniform distribution of sources in the Galactic disc.

Under the latter assumption, the observed positron flux is then given by Eq.  (5.22) of Berezinskii et al. (1990) which can be further simplified to their Eq.  (5.24),
\beq
\frac{dN(E)}{dE}\simeq Q(E) c \frac {V_{\rm{source}} \tau(E) }{V_{\rm{occ}}(E)},
\label{pos_spec}
\eeq
 We have multiplied by c on the r.h.s as our l.h.s has the dimension of 
GeV$^{-1}$ cm$^{-2}$ s$^{-1}$ sr$^{-1}$ and 
  $Q(E)= K E^{-\alpha}$ is the injected positron density in GeV$^{-1}$ cm$^{-3}$ s$^{-1}$ sr$^{-1}$.
The constant $K$ can be calculated for the volume occupied by the sources 
$V_{\rm{source}}$ and the part of the Galaxy occupied by the positrons $V_{\rm{occ}}(E)$. 
In our case, this yields 
\beq
\frac{V_{\rm{source}}}{V_{\rm{occ}}(E)}=\frac{2\pi R^2d}{2\pi R^2\lambda(E)} .
\label{vol_r}
\eeq
We can calculate the injection positron density required to explain the rise in the positron spectrum beyond 30 GeV with eqs. (\ref{pos_spec}) and (\ref{vol_r}). 
With the required injection positron density one can then compute the luminosity in positrons (and thus proton, and gamma-ray)
emission required to explain the observed rise.

\section{Results and discussion}

 We vary the spectral index $\gamma_2$ and the break energy $\epsilon_{\rm{br}}$ to draw a positron spectrum that fits well the rise of the experimental data, see Fig. 1. Assuming  interactions in jets of Doppler factor $\delta_{\rm{D}}=3$, values of $\gamma_2=1.7$, and $\epsilon_{\rm{br}}=0.1$ MeV seem to provide a good fit to the experimental data. Our value of 0.1 MeV is similar to the observed peaks in sources such as GX 339-4 (Hannikainen et al. 1998, Homan et al. 2005) and XTE J1118+480 (McClintock et al. 2001, Zurita et al. 2006, Vila et al. 2012).
 The photon spectrum of spectral index $-1.7$ is produced from synchrotron emissions of electrons and positrons with a spectrum $\frac{dN_{\rm{e}}(E_{\rm{e}})}{dE_{\rm{e}}}\propto E_{\rm{e}}^{-2.4}$. 
This spectrum is similar to the positron spectrum injected from the jets in our model with a spectral index $\gamma_2-4=-2.3$ as discussed in our section 3. Thus our model is self-consistent.

Note that  from the threshold energy condition of pion production in $p\gamma$ 
interactions, if $\epsilon_{\rm{br}}=0.1$ MeV (observers's frame), then the break in the positron spectrum would appear at 630 GeV (also in the observers' frame).

We use our Eq. (\ref{pos_spec}) to calculate $Q(E)$ using the observed positron flux.  At $E=30$ GeV, the observed positron flux is 
$dN(E)/dE=(12/E^3)\times 10^{-4}$ (GeV$^{-1}$ cm$^{-2}$ s$^{-1}$ sr$^{-1}$), see Fig. 1. 
The injection spectrum of positrons is
$Q(E)$ (GeV$^{-1}$ cm$^{-3}$ s$^{-1}$ sr$^{-1}$)=$A/E^{2.3}$ (as for $\gamma_2=1.7$ the injected spectral index into the interstellar medium is ($\gamma_2-4$)). 

With $d=150$ pc, $\lambda(E=30\,{\rm GeV})=600 \, {\rm pc}=1.8\times10^{21}$ cm, and $\tau(E=30\,{\rm GeV})=1/(30\times1.38\times 10^{-16})$ s, we get $Q(E)=6\times 10^{-29}/E^{2.3}$ (GeV$^{-1}$ cm$^{-3}$ s$^{-1}$ sr$^{-1}$). 

The total volume occupied by the sources 
$V_{\rm{s}}=2.8\times 10^{66}$ cm$^{3}$ gives above $E_{\rm{min}}= 30$ GeV the total luminosity in positrons is 
$L_{\rm{pos}}=4\pi V_s \int_{E_{\rm{min}}}^{\infty} E Q(E) dE=4 \times 10^{36}$ erg s$^{-1}$.


Thus to explain the rise in the positron spectrum beyond 30 GeV, the total luminosity injected in positrons from all the LMXBs distributed in the Galactic disc has to be of the order of $ 10^{36}$ erg s$^{-1}$. Note that this is the total luminosity in positrons emitted by all injecting sources in the Galactic disc.
If at least hundreds to thousands of such sources are present in the Milky Way with
typical Eddington's luminosity $\sim 10^{38}$ erg s$^{-1}$ and X-ray luminosity $10^{34}-10^{36}$ erg s$^{-1}$ (Voss \& Ajello 2010), and they emit a 
positron flux with luminosity $\sim 10^{33-34}$ erg s$^{-1}$ each, this population of sources could explain the rise in the observed diffuse positron flux above 30 GeV by itself. 
 Probably, whereas this population contributes to the positron excess, it is not necessarily the only one producing it.

The luminosity distribution functions of the Swift-BAT observed X-ray binaries 
have been derived in Voss \& Ajello 2010. These authors found that X-ray binaries give a local emissivity of $2-4 \times 10^{36}$  erg  s$^{-1}$  Mpc$^{-3}$ to the X-ray background, of which approximately $80\%$ comes from the LMXBs. The incident energy flux from LMXBs in X-rays is $\sim 10^{-11} $ erg  s$^{-1}$ cm$^{-2}$ sr$^{-1}$ which is produced by synchrotron emission from high energy electrons and positrons inside these sources.

As the positrons are losing energy before escaping from their production site in the jet, the emitted luminosity is lower than their original luminosity. If their luminosity at production is ten times higher, then we expect similar luminosity $\sim 10^{34-35}$ erg s$^{-1}$ in gamma-rays and neutrinos from each of these sources. However, distance dilution, beaming, and, in the case of photons, absorption, would make most of these sources individually undetectable. 
 
The neutrinos produced in $p\gamma$ interactions do not lose energy inside or outside the jets unlike positrons. The neutrino spectrum $\phi_{\nu}(E_{\nu})$ (GeV$^{-1}$ s$^{-1}$) has a spectral index $(\gamma_2-3)$ below the break at 630 GeV in the observer's frame and $(\gamma_1-3)$ above, where we assume $\gamma_2=1.7$ and $\gamma_1=1.5$.
At production the flavor ratio of neutrinos in the jet is $\nu_e:\nu_{\mu}:\nu_{\tau}=1:2:0$, subsequently after many oscillations the observed ratio on earth  is expected to be $1:1:1$.
For a jet of luminosity of $3\times10^{35}$erg/sec in neutrinos at a distance of 8 kpc away from us and for maximum energy of protons 20 TeV, the muon neutrino flux expected on earth is calculated below.
The neutrino energy flux is enhanced by a factor of $\delta_D^{(2+x)}$ (assuming the jets are continuous) in the observer's frame (\cite{ghis}) due to Doppler boosting, where $x$ is its spectral index.
The muon neutrino energy flux is expected to be 
$E_{\nu_{\mu}}^2\phi_{\nu_{\mu}}(E_{\nu_{\mu}})= 1.7\times 10^{-10} E_{\nu_{\mu}}^{0.7} $ GeV cm$^{-2}$ s$^{-1}$ and 
$ E_{\nu_{\mu}}^2\phi_{\nu_{\mu}}(E_{\nu_{\mu}})=6\times 10^{-10} E_{\nu_{\mu}}^{0.5}$ GeV cm$^{-2}$ s$^{-1}$ below and above the break energy at 630 GeV, which is much lower than the atmospheric neutrino background. 

A more detailed study in which the sources are not uniformly distributed, and/or they do not present the same injection features would allow to 
constrain whether this explanation may deal with the whole, or part of the positron fraction.
 
\begin{figure}
  \vspace{5mm}
  \centering
  \includegraphics[width=3.in]{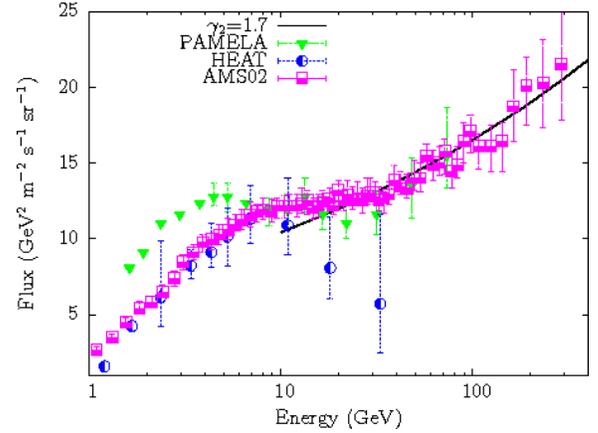}
  \caption{Observed data compared with theoretically obtained positron spectrum in $p\gamma$ interactions in jets of Lorentz factor $\delta_D=3$, solid line: 
$\gamma_2=1.7$, $\epsilon_{br}=0.1$ MeV.}

  \label{pos_flux}
 \end{figure}

\section{Concluding remarks}

 The recent detection of Doppler-shifted X-ray emission lines from a typical black-hole candidate X-ray binary, 4U~1630--47, coincident with the reappearance of radio emission from the jets of the source, implies that baryons can be accelerated in jets of microquasars \citep{diaz}. 
The jets should be strong sources of gamma-rays and neutrinos, and in principle could contribute to the observed positron excess significantly.

The positron excess has been studied earlier with $pp$ interaction models, see e.g., Gaggero et al. (2013), 
and constrained with observed $B/C$ ratios, see, e.g., Cholis \& Hooper (2013), Mertsch \& Sarkar (2014).
Here, we have shown that $p\gamma$ interactions in boosted environments such as jets of microquasars may help in explaining the observed rise in the positron spectrum beyond 30 GeV. 
Low mass microquasars are of special interest in this regard, for hadronic models based on inelastic $pp$ collisions are not expected to play a leading role,
the companion star being cold and old. Because of the same reasons, the external photon background are scarce, what would limit (albeit not rule out in all cases, especially due to self-synchrotron Compton, see, e.g., Bosch-Ramon et al. 2006a,2006b) the ability of leptonic processes to dominate the spectra. If jets accelerate protons, these sources may lead to multi-particle injection via $p\gamma$ processes, where perhaps the photons are synchrotron generated at the base of the jets, see e.g., Levinson \& Waxman 2001.
For recent models of proton low-mass microquasars see e.g., 
Romero \& Vila 2008, 2010; Saitou et al. 2011; Zhang et al. 2010; Vila et al. 2012.
 Luminosities of the $p\gamma$ channel in these works for individual sources
are in agreement with the requirements found in our work in order 
to explain a significant part of the positron excess with microquasar jets.


We finally compare our scenario of $p\gamma$ interactions and subsequent photo-pion decay with the scenario of cosmic ray interactions in the interstellar medium discussed by Cowsik, Burch \& Maziwa-Nussinov (2013). In this paper the authors have interpreted the observed positron spectrum as the secondaries produced in interactions of primary cosmic ray nuclei with interstellar medium assuming the positrons stay for 2 Myr in the Galaxy. According to their prediction the positron spectrum is proportional to $ E_{\rm{e}}^{-3.65}$ above 300 GeV. In our case the spectrum is proportional to $E_{\rm{e}}^{-2.8}$ up to the break energy in the positron spectrum at 630 GeV if the break energy ($\epsilon_{\rm{br}}$) in the synchrotron spectrum of photons from the jets is at 0.1 MeV. Above the break energy the spectral index of the positron spectrum would be proportional to $E_{\rm{e}}^{\gamma_1-4-1/2}$ where $\gamma_1$ is the spectral index of the photon spectrum below $\epsilon_{br}$.   
In some cases, 
the positron flux from these sources could be higher than their electron flux, for instance, if their luminosity in primary protons is higher than in electrons.
If so, we expect these sources to be positron dominated and not contribute significantly to the observed diffuse electron flux.

\section{Acknowledgment}
NG thanks A. Brunnbauer and A. Ibarra for providing the AMS02 data in tabular form. DFT acknowledges the grants AYA2012-39303 and SGR2009-811.

\end{document}